\documentstyle[multicol,aps,prl,epsfig]{revtex}

\renewcommand{\narrowtext}{\begin{multicols}{2} \global\columnwidth20.5pc}
\renewcommand{\widetext}{\end{multicols} \global\columnwidth42.5pc}

\input{epsf.tex}

\begin{document}

\draft

\title{Andreev Reflection in Strong Magnetic Fields}

\author{H.~Hoppe, U.~Z\"ulicke, and Gerd Sch\"on}

\address{Institut f\"ur Theoretische Festk\"orperphysik,
Universit\"at Karlsruhe, D-76128 Karlsruhe, Germany}

\date{submitted to Phys.~Rev.~Lett. 26 July 1999;
revised 8 November 1999}

\maketitle

\begin{abstract}
We have studied the interplay of Andreev reflection and
cyclotron motion of quasiparticles at a
superconductor--normal-metal interface with a strong magnetic
field applied parallel to the interface. Bound states are
formed due to the confinement introduced both by the external
magnetic field and the superconducting gap. These bound states
are a coherent superposition of electron and hole edge
excitations similar to those realized in finite
quantum-Hall samples. We find the energy spectrum for these
Andreev edge states and calculate transport properties.
\end{abstract}


\narrowtext

Rapid progress in fabrication techniques has made it possible to
investigate phase-coherent transport in a variety of mesoscopic
conducting devices~\cite{mesoeltrans}. In recent years, the
study of hybrid systems consisting of superconductors in contact
with normal metals has continued to attract considerable
interest, mainly because of the novel effects observed in
superconductor--semiconductor microjunctions~\cite{reviews}. Many
of the unusual experimental findings arise due to the phenomenon
of {\em Andreev reflection\/}, i.e., the (partial)
retroreflection of an electron incident on a
superconductor~(S)~--~normal-metal~(N) interface as a
hole~\cite{andreev,btk}. Phase coherence between the electron
and hole states is maintained during the reflection process.
Hence, coupled-electron-hole (Andreev) bound
states~\cite{andreev} having energies within the superconducting
gap are formed in mesoscopic devices with multiple interfaces,
e.g., S--N--S systems~\cite{kulik}, or S--N--I--N--S
structures~\cite{adz:jetp:80}. (The symbol `I' denotes an
insulating barrier.) Recently, measurements of transport across
the interface between a superconductor and a two-dimensional
electron gas (2DEG) were performed with a strong magnetic field
$H$ applied in the direction perpendicular to the plane of the
2DEG~\cite{expt}. While the magnetic field did not exceed the
critical field of the superconductor, it was still large enough
such that the Landau-level quantization of the electronic motion
in the 2DEG was important~\cite{lowfield}. With these experiments,
a link has been established between mesoscopic superconductivity
and quantum-Hall physics~\cite{qhe-sg} which needs theoretical
exploration.

In this Letter, we study a novel kind of Andreev bound state that
is formed {\em at a single S--N  interface\/} in a strong
magnetic field~\cite{previous}. This bound state is a coherent
superposition of an electron and a hole propagating along the
interface in a new type of current-carrying edge state that is
induced by the superconducting pair potential. Andreev reflection 
gives rise to the contribution
\begin{equation}\label{condres}
G_{\text{AR}}=\frac{e^2}{\pi\hbar}\,\sum_{n=1}^{n^*} B_n
\end{equation}
to the small-bias conductance, which we obtained by generalizing
the familiar B\"uttiker description~\cite{butt:prb:88} of
transport in quantum-Hall samples. In Eq.~(\ref{condres}), the
summation is over Andreev-bound-state levels that intersect with
the chemical potential, and $B_n$ is the hole probability for a
particular bound-state level. It turns out that $n^*$ is twice
the number of orbital Landau levels occupied in the bulk of the
2DEG, and $B_n\le 1/2$ depends weakly on magnetic field $H$ for
an ideal interface but oscillates strongly with $H$ for a
non-ideal interface. $G_{\text{AR}}$ can be measured directly as
the two-terminal conductance in a S--2DEG--S system~\cite{expt}.
Our treatment in terms of Andreev edge states provides a clear
physical description of transport in such devices and explains
oscillatory features in the conductance that were observed
experimentally~\cite{expt} and also obtained in previous numerical
studies~\cite{ytak:prb:98a}.

\begin{figure}
\centerline{\epsfig{file=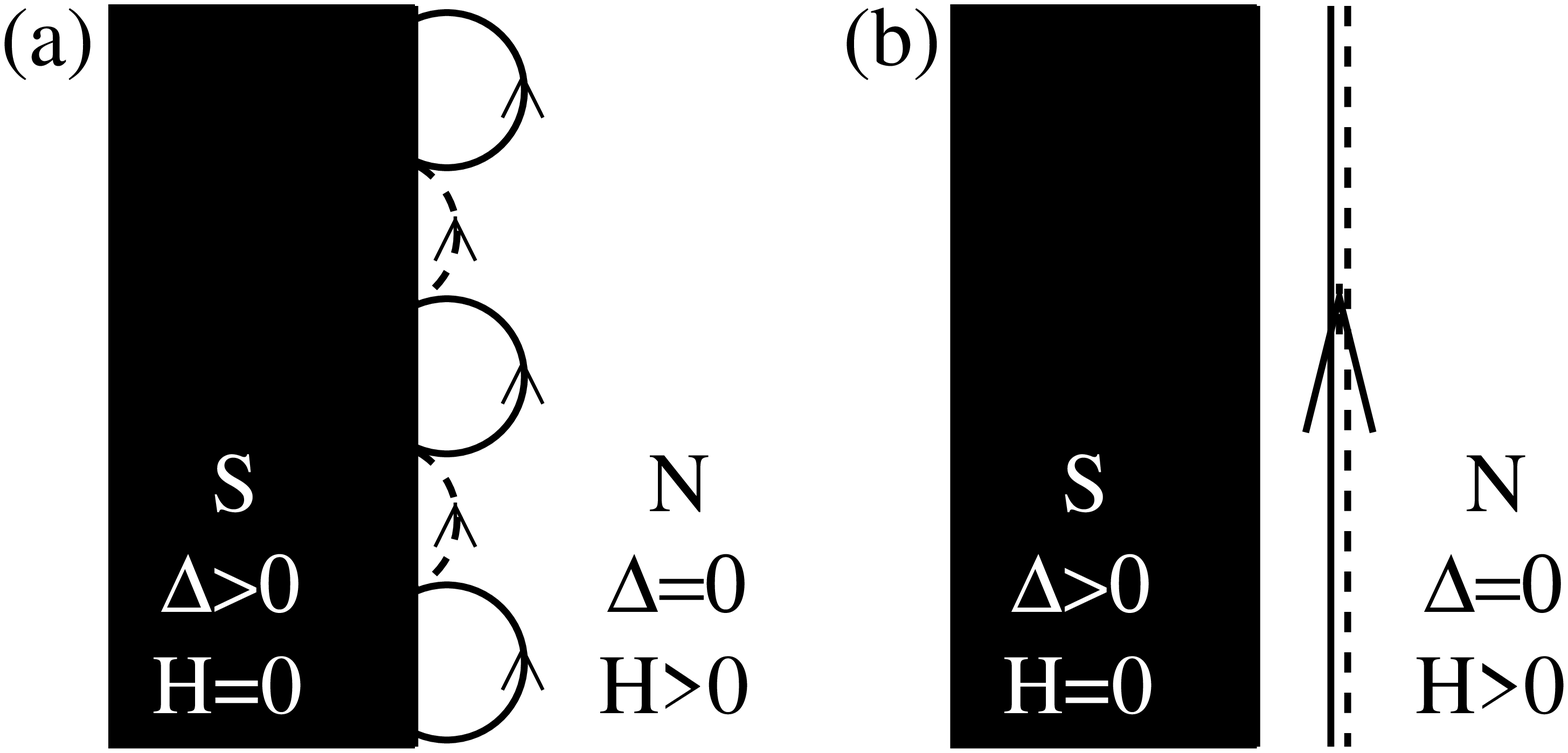,width=2.7in}}
\vspace{0.3cm}
\caption{Andreev bound state of an electron (solid lines) and a
hole (dashed lines) at an S--N interface in a magnetic field.
(a)~Classical picture of electron and hole alternating in
skipping orbits. (b)~The quantum-mechanical analysis finds that
both the electron and the hole occupy Landau-level states that
are extended parallel to the interface.}
\label{skipquant}
\end{figure}

Let us start by recalling the classical and quantum-mechanical
descriptions of electron dynamics in an external magnetic field.
When considered to be classical charged particles, bulk-metal
electrons execute periodic cyclotron motion with a frequency
$\omega_{\text{c}}=eH/(m c)$. A surface that is parallel to the
magnetic field interrupts the cyclotron orbits of nearby electrons
and forces them to move in skipping orbits along the
surface~\cite{prange}. Within the more adequate quantum-mechanical
treatment, the kinetic energy for electronic motion in the plane
perpendicular to the magnetic field is quantized in Landau
levels~\cite{ldl:zphys:30} which are at constant eigenvalues
$\hbar\omega_{\text{c}}(n+1/2)$ for electron states localized in
the bulk but are bent upward in energy for states localized close
to the surface~\cite{bih:prb:82}. Applying the classical picture
of cyclotron and skipping orbits to a S--N
interface~\cite{negcyc}, one finds that Andreev reflection leads
to electrons and holes alternating in skipping orbits along the
interface. [See Fig.~\ref{skipquant}(a).] In what follows, we
provide a full quantum-mechanical description of these alternating
skipping orbits in terms of current-carrying Andreev bound states.
[See Fig.~\ref{skipquant}(b).]

We now provide details of our calculation. A planar interface is
considered, located at $x=0$ between a semi-infinite region
($x<0$) occupied by a type-I superconductor and a semi-infinite
normal region ($x>0$). A uniform magnetic field is applied in $z$
direction, which is screened from the superconducting region due
to the Meissner effect. Neglecting inhomogeneities in the
magnetic field due to the existence of a finite penetration
layer~\cite{meissner}, we assume an abrupt change of the
magnetic-field strength at the interface: $H(x)=H\,\Theta(x)$,
where $\Theta(x)$ is Heaviside's step function. The energy
spectrum of Andreev bound states at the S--N interface is found
by solving the Bogoliubov--de~Gennes (BdG)
equation~\cite{degennes},
\begin{equation}\label{bdg}
\left(\begin{array}{cc} H_{\text{0},+} + U_{\text{ext}} &
\Delta \\ \Delta^{\text{*}} & - H_{\text{0},-} - U_{\text{ext}}
\end{array}\right)\left( \begin{array}{c} u \\ v \end{array}
\right)= E\, \left( \begin{array}{c} u \\ v \end{array}\right)
\, ,
\end{equation}
with spatially non-uniform single-electron/hole Hamiltonians
$H_{\text{0},\pm}$ and pair potential~\cite{pairapology}
$\Delta(x)=\Delta_{\text{0}}\,\Theta(-x)$. We introduced the
potential $U_{\text{ext}}(x)=U_{\text{0}}\,\delta(x)$ to model
scattering at the interface, and allow the effective mass and
the Fermi energy to be different in the superconducting and
normal regions. Choosing the vector potential $\vec A(x) = x\, H
\,\Theta(x)\,\hat y$, we have
\begin{equation}\label{hamilt}
H_{\text{0},\pm} = \left\{ \begin{array}{cl}\frac{p_x^2+p_z^2}
{2 m_{\text{N}}}+\frac{m_{\text{N}}\omega_{\text{c}}^2}{2}\left(
x \mp X_{p_y} \right)^2 - \epsilon_{\text{F}}^{\text{(N)}} & x>0
\\ \frac{p_x^2+p_y^2+p_z^2}{2 m_{\text{S}}}-
\epsilon_{\text{F}}^{\text{(S)}} & x<0 \end{array}\right. .
\end{equation}
The operator $X_{p_y}=p_y\ell^2\, {\mathrm sgn}(e H)/\hbar$ is
the guiding-center coordinate in $x$ direction for cyclotron
motion of electrons in the normal region, and $\ell= \sqrt{\hbar
c/|e H|}$ denotes the magnetic length. Uniformity in the $y$ and
$z$ directions suggests the separation {\it ansatz\/}
\begin{mathletters}\label{sepansatz}
\begin{eqnarray}
u(x,y,z) &=& f_X(x)\,\,e^{i y\, X/\ell^2}\,e^{i k z}\,/\sqrt{L_y
\, L_z}\,\, ,\\
v(x,y,z) &=& g_X(x)\,\,e^{i y\, X/\ell^2}\,e^{i k z}\,/\sqrt{L_y
\, L_z}\,\, .
\end{eqnarray}
\end{mathletters}
The lengths $L_y$, $L_z$ are the sample sizes in $y$ and $z$
directions, respectively. Solutions of Eq.~(\ref{bdg}) for the
S--N junction are found by matching appropriate wave functions
that are solutions in the normal and superconducting regions,
respectively~\cite{btk}. The motion in $z$ direction can
trivially be accounted for by renormalized Fermi energies $\tilde
\epsilon_{\text{F}}^{\text{(N,S)}}=\epsilon_{\text{F}}^{
\text{(N,S)}}-\hbar^2 k^2/(2 m_{\text{N,S}})$. Non-trivial
matching conditions arise only in $x$ direction. For fixed $X$
and $|E|<\Delta_{\text{0}}$, we have to match at $x=0$ the wave
function
\begin{mathletters}
\begin{equation}\label{normalside}
\left(\begin{array}{c} f_X\\g_X\end{array}\right)_{x>0} = 
\left(\begin{array}{c} a\,\,\chi_{\varepsilon_+}(x - X) \\ b\,\,
\chi_{\varepsilon_-}(x+X)\end{array}\right)\, ,
\end{equation}
corresponding to a coherent superposition of an electron and a
hole in the normal region, to that of evanescent excitations in
the superconductor,
\begin{equation}
\left(\begin{array}{c} f_X\\g_X\end{array}\right)_{x<0} = 
d_+\left(\begin{array}{c}\gamma\\1\end{array}
\right)e^{i\, x \lambda_-} + d_-\left(\begin{array}{c}\gamma^*\\1
\end{array}\right) e^{-i\, x \lambda_+} \, .
\end{equation}
\end{mathletters}
The parameters $\gamma$ and $\lambda_\pm$ are defined in the usual
way~\cite{kulik}. The functions $\chi_{\varepsilon_\pm}(\xi)$
solve the familiar one-dimensional harmonic-oscillator
Schr\"odinger equation,
\begin{equation}\label{parboleq}
\frac{\ell^2}{2}\,\frac{d^2\chi_{\varepsilon_\pm}}{d\xi^2}-
\left[\frac{\xi^2}{2\ell^2}-\frac{\varepsilon_\pm}{\hbar
\omega_{\text{c}}}\right]\,\chi_{\varepsilon_\pm} = 0\quad ,
\end{equation}
with eigenvalues $\varepsilon_\pm=\epsilon_{\text{F}}^{\text{(N)}
}\pm E-\hbar^2 k^2/(2m_{\text{N}})$ and are assumed to be
normalized to unity in the normal region. Hence, they are
proportional to the fundamental solutions of Eq.~(\ref{parboleq})
that are well-behaved for $x\to\infty$; these are the
{\em parabolic cylinder functions\/}~\cite{abramowitz}
$U(-\frac{\varepsilon_\pm}{\hbar \omega_{\text{c}}}, 
\frac{\sqrt{2}\xi}{\ell})$. The matching conditions yield a
homogeneous system of four linear equations for the coefficients
$a,b,d_\pm$ whose secular equation determines the allowed values
of $E$. 

It is straightforward to calculate the probability and charge
currents for any particular Andreev-bound-state solution of the
BdG equation~(\ref{bdg}) that is labeled by guiding-center
coordinate $X$ and energy $E$. It turns out that currents flow
parallel to the interface. The total (integrated) quasiparticle
{\em probability\/} current is given by
\begin{equation}
I_X^{\text{(P)}} = \frac{1}{\hbar}\,\frac{\ell^2}{L_y}\,
\frac{\partial E}{\partial X}\quad . 
\end{equation}
The total {\em charge\/} current can be written as the sum of
three contributions, $I_X^{\text{(Q)}} = I_X^{\text{(Q,n)}} -
I_X^{\text{(Q,a)}} + I_X^{\text{(Q,s)}}$, where
\begin{mathletters}\label{currents}
\begin{eqnarray}
I_X^{\text{(Q,n)}} &=& \frac{e}{\hbar}\,\frac{\ell^2}{L_y}\,
\frac{\partial E}{\partial X} \quad , \\
I_X^{\text{(Q,a)}} &=& \frac{e}{\hbar}\,\frac{\ell^2}{L_y}\,
\frac{\partial E}{\partial X}\,\, 2\int_x\, \left|\, g_X(x)\,
\right|^2 \quad ,\\
I_X^{\text{(Q,s)}} &=& \frac{e}{\hbar}\,\frac{\ell^2}{L_y}\,
2\, \Delta\,\,\int_x \, \Theta(-x) \left[g_X^*\frac{d f_X}{d X} -
f_X^*\frac{d g_X}{d X} \right]\, .
\end{eqnarray}
\end{mathletters}
Note that $I_X^{\text{(Q,n)}}$ is the current that would flow
in an ordinary quantum-Hall edge state~\cite{bih:prb:82}, i.e.,
due to normal reflection at the interface. The existence of
Andreev reflection is manifested in the contribution
$-I_X^{\text{(Q,a)}}$ to the Hall current; it is proportional to
the hole probability $B(X) = \int_x\, |\, g_X(x)\,|^2$. The part
$I_X^{\text{(Q,s)}}$ of the quasiparticle charge current is
converted into a supercurrent.

Numerical implementation of the matching procedure is
straightforward. More detailed insight is gained, however, when
considering the limit $|X|\ll\sqrt{\varepsilon_\pm/(2m_{\text{N}}
\omega_{\text{c}}^2)}$ for which analytical progress can be made.
Using an asymptotic form for the parabolic cylinder
functions~\cite{abramowitz}, the secular equation can be written
as
\begin{equation}\label{secular}
\cos(\varphi_+) + \Gamma(\varphi_-) = \frac{2 s}{s^2+w^2+1}\,
\frac{E\,\sin(\varphi_+)}{\sqrt{\Delta_{\text{0}}^2 - E^2}}\quad .
\end{equation}
Here we used the Andreev approximation~\cite{andreev} ($E,
\Delta_{\text{0}}\ll\tilde\epsilon_{\text{F}}^{\text{(N)}},\tilde
\epsilon_{\text{F}}^{\text{(S)}}$), and the abbreviations
\begin{mathletters}
\begin{eqnarray}
\varphi_+ &=& \pi\frac{E}{\hbar\omega_{\text{c}}} + \frac{2 X}
{\hbar}\sqrt{2m_{\text{N}}\tilde\epsilon^{\text{(N)}}_{\text{F}}}
\quad , \\
\varphi_- &=& \pi\,\frac{\nu}{2} + \frac{E\,X}{\hbar}\,
\sqrt{\frac{2m_{\text{N}}}{\tilde\epsilon^{\text{(N)}}_{\text{F}}}}
\quad , \\
\Gamma(\alpha) &=& \frac{[s^2+w^2-1]\sin(\alpha)+2 w \cos(\alpha)}
{s^2+w^2+1}\quad .
\end{eqnarray}
\end{mathletters}
The variable $\nu=2\,\tilde\epsilon_{\text{F}}^{\text{(N)}}/(\hbar
\omega_{\text{c}})$ coincides with the {\em filling factor\/} of
quantum-Hall physics~\cite{qhe-sg} when the N region is a 2DEG. The
parameter $s=[\tilde\epsilon_{\text{F}}^{\text{(S)}}m_{\text{N}}/
(\tilde\epsilon_{\text{F}}^{\text{(N)}} m_{\text{S}})]^{1/2}$
measures the Fermi-velocitiy mismatch for the junction, and
$w=[2m_{\text{N}}U_{\text{0}}^2/(\hbar^2\tilde
\epsilon_{\text{F}}^{\text{(N)}})]^{1/2}$ quantifies interface
scattering. We discuss briefly results for two limiting
cases~\cite{details}.

{\it (a)~Ideal interface.\/} In the absence of scattering at the
S--N interface ($w=0$) and for perfectly matching Fermi velocities
($s=1$), $\Gamma(\alpha)$ vanishes identically. The energy
spectrum is found from solutions of the transcendental equation
$\cot(\varphi_+)=E/\sqrt{\Delta_{\text{0}}^2 - E^2}$. It consists
of several bands, and states within each band are labeled by their
guiding-center quantum number $X$. It turns out that $a^2=b^2$
{\em at any energy\/}, and the band dispersion is
\begin{equation}
\frac{\partial E}{\partial X} = -\frac{2\sqrt{2m_{\text{N}}\tilde
\epsilon^{\text{(N)}}_{\text{F}}}}{\hbar}
\frac{\sqrt{\Delta_{\text{0}}^2 - E^2}}{1 + \pi
\sqrt{\Delta_{\text{0}}^2 - E^2}/(\hbar\omega_{\text{c}})}
\quad .
\end{equation}

{\it (b)~Non-ideal (S--I--N) interface at low energies.\/}
For $E\ll{\mathrm min}\{\Delta_{\text{0}},\hbar\sqrt{
\tilde\epsilon^{\text{(N)}}_{\text{F}}/(2 m_{\text{N}} X^2)}\}$, the
dependence of $\varphi_-$ on $E$ and $X$ can be neglected. We find
\begin{equation}\label{nonideal}
E=\Delta_{\text{0}}\,\,\frac{(2n+1)\pi\pm{\mathrm arccos}(
\Gamma_{\text{0}}) - 2X\sqrt{2 m_{\text{N}}\tilde
\epsilon^{\text{(N)}}_{\text{F}}}/\hbar}{q+\pi\Delta_{\text{0}}/
(\hbar\omega_{\text{c}})}\, ,
\end{equation}
where $\Gamma_{\text{0}}=\Gamma(\pi\nu/2)$ and $q=2s/(s^2+w^2+1)$.
For $s=1$ and $w=0$, the spectrum for an ideal interface at small
energies emerges. In the opposite limit of a very bad interface
($s^2+w^2\to\infty$), we recover the spectrum of the Landau-level
edge states close to a hard wall~\cite{bih:prb:82}. In general,
$\Gamma_{\text{0}}$ oscillates as a function of filling factor
$\nu$. Hence, unlike in the ideal case, the bound-state energies
of Eq.~(\ref{nonideal}) vary oscillatory with $\nu$. The same is
true for the hole probability $B\approx b^2\le 1/2$, for which we
find
\begin{equation}\label{nonidhole}
B = \frac{1}{2}\,\,\frac{q^2/(1-\Gamma_{\text{0}}^2)}{1+\sqrt{1-
q^2/(1-\Gamma_{\text{0}}^2)}}\quad .
\end{equation}
The minima in the oscillatory dependence of $B$ on filling factor
$\nu$ occur whenever $\tan(\pi\nu/2) = 2w/(1-s^2-w^2)$.

\begin{figure}
\centerline{\epsfig{file=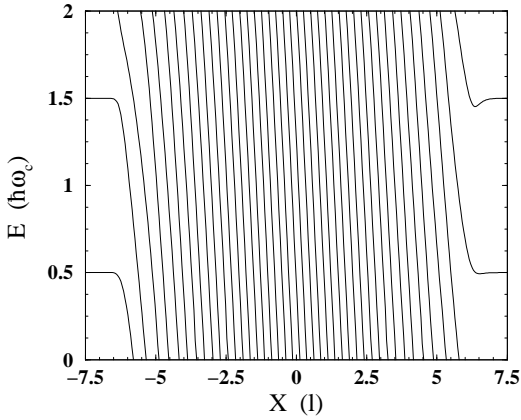,width=2.7in}}
\caption{Andreev-bound-state levels for the ideal S--N interface
in a magnetic field, calculated numerically for $\nu=40$ and
$\Delta_{\text{0}}/(\hbar\omega_{\text{c}})=2$ by exactly
matching solutions of the BdG equation for the normal and
superconducting regions.}
\label{bentLL}
\end{figure}

Results obtained in the approximate analytical treatment sketched
above are expected to be valid only as long as $X$ is not too
large. It turns out, however, that they actually provide a good
description at $E\approx 0$ for Andreev levels intersecting with
the Fermi energy, which are important for small-bias transport. In
particular, we obtained a non-vanishing dispersion $\partial E/
\partial X$ close to the interface which leads to a finite Hall
current $I_X^{\text{(Q,n)}}-I_X^{\text{(Q,a)}}$.
[See Eqs.~(\ref{currents}).] It is clear that,
far away from the interface, i.e., for $|X|\gg r_{\text{c}}$, no
coupling of electrons and holes via the pair potential is possible
and dispersionless Landau levels are solutions of the BdG
equation. However, as the guiding-center coordinate $X$ gets close
to the interface, these Landau levels are bent upward and become
Andreev-bound-state levels for $|E|<\Delta_{\text{0}}$. This is
seen in the exact numerical calculation of the spectrum
(Fig.~\ref{bentLL}), which also provides a crucial piece of
information that is elusive within the approximate analytical
treatment: the number $n^*$ of Andreev levels intersecting with
the Fermi energy. We find that $n^*$ is twice the integer part of
$\nu/2$.

\begin{figure}
\centerline{\epsfig{file=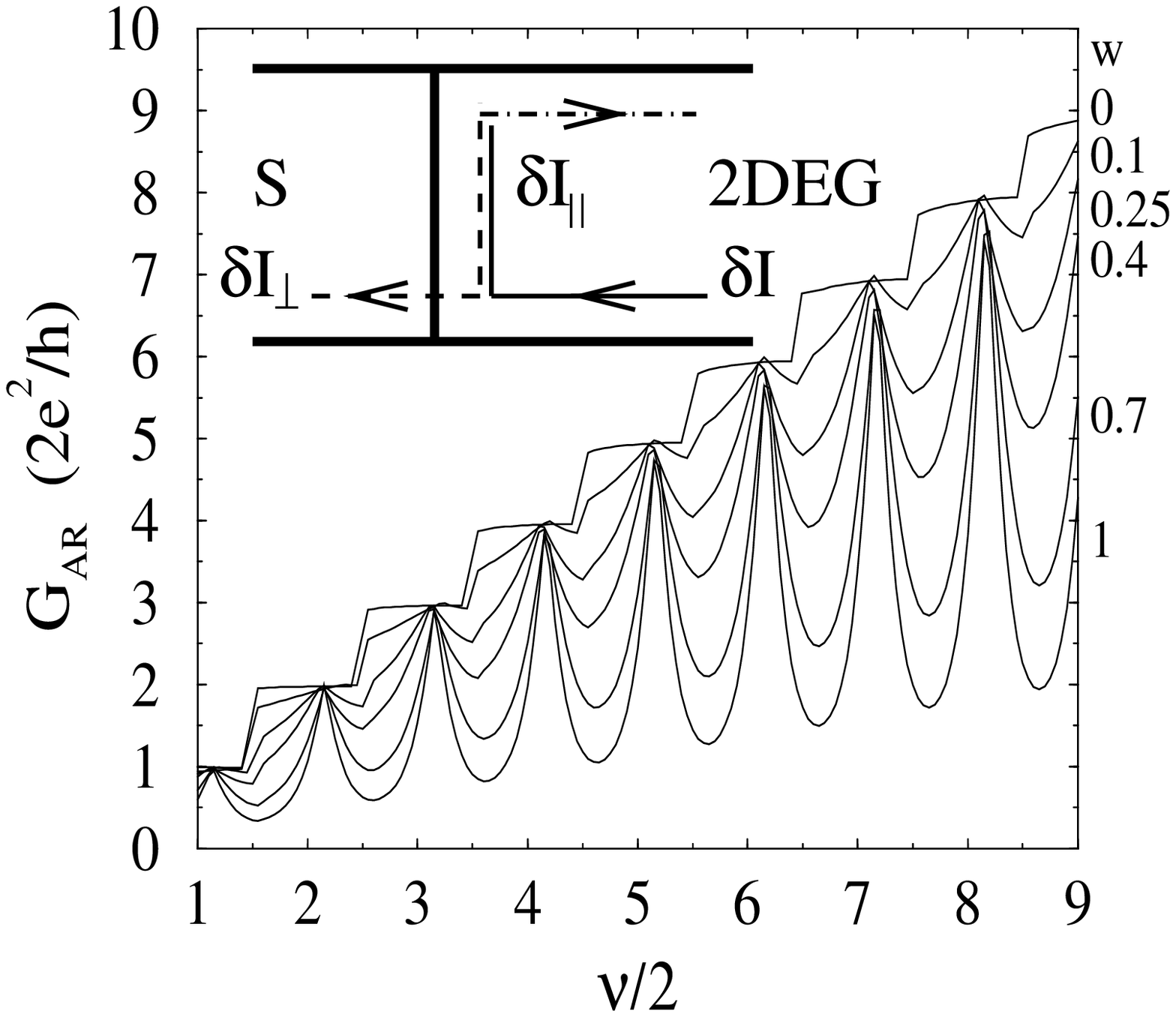,width=2.9in}}
\caption{Conductance $G_{\text{AR}}$, calculated according to
Eq.~(\ref{condres}) with hole probabilities $B_n$ obtained from
the exact numerical matching procedure. We set $\Delta_0=
0.02~\tilde\epsilon_{\text{F}}^{\text{(N)}}$ and $s=1$. The value
of $w$ for each curve is indicated. The inset shows how the current
$\delta I$ carried by the lower quantum-Hall edge channel (solid 
line) is distributed at the left interface. The part $\delta
I_\parallel$ flows in Andreev edge states; it has both normal
(solid line) and Andreev-reflection (dashed line) contributions.
It is collected by the upper quantum-Hall edge channel (dot-dashed
line) and returned to the right interface. As quasiparticle
probability current is conserved, and backscattering prohibited
due to chirality, $\delta I_\perp$ equals the Andreev-reflection
contribution to $\delta I_\parallel$ (dashed line).}
\label{transpfig}
\end{figure}

We now apply our findings to study transport in S--2DEG--S
structures~\cite{expt}. In experiment, two S--N interfaces are
linked by ordinary quantum-Hall edge channels whose local chemical
potentials differ by $\delta\mu$. Generalizing the B\"uttiker
formalism~\cite{butt:prb:88} for edge-channel transport, the
following picture emerges. (See inset of Fig.~\ref{transpfig}.)
From the lower edge channel, a current
$\delta I=\delta\mu\cdot e/h$ impinges on the left S--N interface.
This current divides up into a part $\delta I_\parallel$ flowing
parallel to the interface in Andreev edge states studied above,
and $\delta I_\perp$ which flows across the interface. Chirality
of edge states (both Andreev and ordinary) and conservation of
quasiparticle probability current completely determines the
current parallel to the interface to be $\delta I_\parallel=
(1-2 B_n)\,\delta\mu\cdot e/h$. The upper edge channel collects
$\delta I_\parallel$ and returns to the right interface, where a
similar discussion applies. Hence, the two-terminal conductance
$e\,\delta I_\perp/\delta\mu$ in the S--2DEG--S device equals
$G_{\text{AR}}$ [given in Eq.~(\ref{condres})]. Using hole
amplitudes obtained from the exact numerical matching procedure,
we determined the filling-factor dependence of $G_{\text{AR}}$
(shown in Fig.~\ref{transpfig} for $2\le\nu\le 18$ and various
values of $w$). As anticipated from the approximate analytical
result $B_n\approx 1/2$, the ideal interface exhibits almost
perfect conductance steps in units of $2e^2/h$. For finite
scattering at the interface, oscillations appear in the
conductance whose amplitude increases as the interface quality
worsens. However, for certain single values of $\nu$, the ideal
conductance is reached even at a bad interface. The location of
minima and maxima in the field dependence of the conductance can
be obtained from our analytical calculation and compare well with
results of previous numerical studies~\cite{ytak:prb:98a} based on
a representation in terms of scattering states.

We thank W.~Belzig, C.~Bruder, T.~M.~Klapwijk, A.~H.~MacDonald,
and A.~D.~Zaikin for useful discussions, and 
Sonderforschungsbereich 195 der DFG for support.

\newpage

\widetext


\begin{thebibliography}{10}

\bibitem{mesoeltrans}
{\em Mesoscopic Electron Transport}, Vol.~345 of {\em NATO ASI
Series E}, edited by L.~L. Sohn, L.~P. Kouwenhoven, and G.
Sch\"on (Kluwer Academic,  Dordrecht, 1997).

\bibitem{reviews}
For reviews and extensive lists of references, see C.~W.~J.
Beenakker, in {\em Mesoscopic Quantum Physics}, {\em Proceedings
of the 1994 Les Houches Summer School, Session LXI}, edited by
E. Akkermans {\it et al.\/} (Elsevier Science, Amsterdam, 1995),
pp.\ 279--324; B.~J. van~Wees and H. Takayanagi, in
Ref.~\onlinecite{mesoeltrans}, pp.\ 469--501; C.~J. Lambert and
R. Raimondi, J. Phys.: Condens. Matter {\bf 10},  901  (1998).

\bibitem{andreev}
A.~F. Andreev, Zh. Eksp. Teor. Fiz. {\bf 46},  1823  (1964);
{\bf 49},  655  (1965) [Sov. Phys. JETP  {\bf 19},
1228 (1964); {\bf 22}, 455 (1966)].

\bibitem{btk}
G.~E. Blonder, M. Tinkham, and T.~M. Klapwijk, Phys. Rev. B
{\bf 25},  4515  (1982).

\bibitem{kulik}
I.~O. Kulik, Zh. Eksp. Teor. Fiz. {\bf 57},  1745  (1969) [Sov.
Phys. JETP {\bf 30}, 944 (1970)].

\bibitem{adz:jetp:80}
A.~D. Zaikin and G.~F. Zharkov, Zh. Eksp. Teor. Fiz. {\bf 78}, 
721  (1978) [Sov. Phys. JETP {\bf 51}, 364 (1980)].

\bibitem{expt}
H. Takayanagi and T. Akazaki, Physica B {\bf 249-251},  462
(1998); T.~D. Moore and D.~A. Williams, Phys. Rev. B {\bf 59}, 
7308 (1999).

\bibitem{lowfield}
Previous studies of the magnetic-field dependence of Andreev
reflection in S--2DEG--S systems were limited to the low-field
regime. See, e.g., J. Nitta, T. Akazaki, and H. Takayanagi, Phys.
Rev. B {\bf 49}, 3659 (1994); J.~P. Heida, B.~J. van~Wees, T.~M.
Klapwijk, and G. Borghs, Phys. Rev. B {\bf 57}, R5618 (1998);
and references therein.

\bibitem{qhe-sg}
{\em The Quantum Hall Effect}, edited by R.~E. Prange and S.~M.
Girvin, 2nd ed.  (Springer, New York, 1990).

\bibitem{previous}
The coupling of a quantum-Hall system to superconducting leads
via tunnel barriers was considered previously. See M. Ma and
A.~Yu. Zyuzin, Europhys. Lett. {\bf 21}, 941 (1993); M.~P.~A.
Fisher, Phys. Rev. B {\bf 49}, 14550 (1994); Y. Ishikawa and H.
Fukuyama, J. Phys. Soc. Jpn. {\bf 68}, 954 (1999). In our work,
we provide a theory of Andreev reflection for arbitrary
transmission of the interface which is similar in spirit to that
given by Blonder, Tinkham, and Klapwijk~\cite{btk} for the
zero-field case.

\bibitem{butt:prb:88}
M. B\"uttiker, Phys. Rev. B {\bf 38},  9375  (1988).

\bibitem{ytak:prb:98a}
Y. Takagaki, Phys. Rev. B {\bf 57},  4009  (1998); Y. Asano,
cond-mat/9909118.

\bibitem{prange}
T.-W. Nee and R.~E. Prange, Phys. Lett. {\bf 25A},  582  (1967);
R.~E. Prange and T.-W. Nee, Phys. Rev. {\bf 168},  779  (1968).

\bibitem{ldl:zphys:30}
L.~D. Landau, Z. Phys. {\bf 64},  629  (1930).

\bibitem{bih:prb:82}
B.~I. Halperin, Phys. Rev. B {\bf 25},  2185  (1982);
A.~H. MacDonald and P. St{\v r}eda, {\it ibid.\/} {\bf 29},  1616
(1984).

\bibitem{negcyc}
In certain mesoscopic systems where electron trajectories extend
only over distances that are much smaller than the cyclotron
radius $r_{\text{c}}=v_{\text{F}}/\omega_{\text{c}}$, a simplified
treatment of magnetic-field effects (neglecting cyclotron motion
altogether) applies. Previous studies of the effect of a magnetic
field on the Andreev-bound-state spectrum in S--N--S and S--N--I
structures used this approximation. See, e.g.,
V.~P. Gala{\u\i}ko, Zh. Eksp. Teor. Fiz. {\bf 57},  941  (1969)
[Sov. Phys. JETP {\bf 30}, 514 (1970)]; V.~P. Gala{\u\i}ko and
E.~V. Bezugly\u\i, {\it ibid.\/} {\bf 60},  1471 (1971)
[{\it ibid.\/} {\bf 33}, 796 (1971)]; G.~A. Gogadze and I.~O.
Kulik, {\it ibid.\/} {\bf 60},  1819  (1971) [{\it ibid.\/}
{\bf 33}, 984 (1971)]; and Ref.~\cite{adz:jetp:80}.  
Here we consider the problem of Andreev reflection in a
{\em strong\/} magnetic field and are therefore in the opposite
limit.

\bibitem{meissner}
This is a simplifying assumption which can be improved upon.
However, we
do not expect large quantitative corrections to the results
obtained within the simple model.

\bibitem{degennes}
P.~G. de~Gennes, {\em Superconductivity of Metals and Alloys}
(Addison-Wesley, Reading, MA, 1989).

\bibitem{pairapology}
We adopt the usual model~\cite{btk} where self-consistency of
the superconducting pair potential is not enforced.

\bibitem{abramowitz}
M. Abramowitz and I.~A. Stegun, {\em Handbook of Mathematical
Functions} (Dover Publications, New York, 1972).

\bibitem{details}
Details of the approximate analytical calculations will be
given in a later publication.

\end{thebibliography}
\end{document}